# Quantum Relaxation for Linear Systems in Finite Element Analysis

Osama Muhammad Raisuddin[1]*, Suvranu De[2]

*raisuo@rpi.edu

[1] *Mechanical, Aerospace, and Nuclear Engineering, Rensselaer Polytechnic Institute, 110 8th St. Troy, NY, 12180, USA*

[2] *Florida Agricultural and Mechanical University-Florida State University College of Engineering, 2525 Pottsdamer St., Tallahassee, FL 32310, USA*

## Abstract

Quantum linear system algorithms (QLSAs) for gate-based quantum computing can provide exponential speedups for solving linear systems but face challenges when applied to finite element problems due to the growth of the condition number with problem size. Furthermore, QLSAs cannot use an approximate solution or initial guess to output an improved solution. Here, we present Quantum Relaxation for Linear System (qRLS), as an iterative approach for gate-based quantum computers by embedding linear stationary iterations into a larger block linear system. The condition number of the block linear system scales linearly with the number of iterations independent of the size and condition number of the original system. The well-conditioned system enables a practical iterative solution of finite element problems using the state-of-the-art Quantum Signal Processing (QSP) variant of QLSAs, for which we provide numerical results using a quantum computer simulator. The iteration complexity demonstrates favorable scaling relative to classical architectures, as the solution time is independent of system size and requires O(log(N)) qubits. This represents an exponential efficiency gain, offering a new approach for iterative finite element problem-solving on quantum hardware.

## 1. Introduction

The solution of linear systems of equations is central to solving physical problems discretized using finite element methods. Classical solution approaches for linear systems scale at best linearly with $N$, the number of unknowns, while quantum linear system algorithms (QLSAs) scale exponentially better as $O(\log N)$ [1]. As an example, the conjugate gradient algorithm scales as $O(N\kappa \log(1/\epsilon))$, where $\kappa$ is the condition number of the system and $\epsilon$ is the desired precision. For the same system of equations, the quantum signal processing algorithm scales as $O(\kappa \log(N\kappa/\epsilon))$ [2]. Although the scaling with $N$ is favorable, the condition number $\kappa$ prevents speedups for finite element problems due to implicit dependence on $N$ [3]. In this paper, we propose an iterative approach to reduce the dependence of the solution complexity on the condition number.

Quantum computing is an emerging computational paradigm with immense potential for speedups compared to classical computing techniques. Quantum computers exhibit the properties of superposition, entanglement and interference in exponentially large state spaces, distinguishing them from classical computers. Quantum computers use quantum bits, 'qubits', to represent information. The state of an $n$-qubit quantum memory register is represented as an exponentially large vector in $\mathcal{H}^{2^n}$. Qubits can exist in a continuum of superposition states of basis vectors. Individual qubits can be 'entangled', which makes their states correlated to each other. Superpositions of states can be made to interfere with one another to cancel out unwanted states. Conceived initially to simulate quantum mechanical systems [4], the application of quantum algorithms extends to a plethora of problems, including factoring large integers [5]. Among these algorithms, the most promising ones for scientific and engineering computation are quantum algorithms for linear systems of equations [1], systems of ordinary differential equations [6], and partial differential equations [7], all of which can potentially provide exponential speedups.

The two prevalent quantum computing architectures are gate-based quantum computing [8] and quantum annealing [9]. Gate-based quantum computing applies discrete operations known as 'quantum gates' on quantum bits or qubits to form a quantum circuit. The output of the quantum circuit can be in the form of a quantum state or measurements of the quantum state. Quantum annealing, a subset of adiabatic quantum computing, is a continuous transition from an initial Hamiltonian to a final Hamiltonian applied to a system of qubits. The adiabatic theorem [10] implies that if the system of qubits is initially in a known ground state of the initial Hamiltonian, the final state of the qubits will be a ground state of the final Hamiltonian if the transition is slow enough.

Quantum computers are error-prone in the presence of noise and imperfections in device fabrication or control hardware. Current quantum devices are dubbed Noisy Intermediate-Scale Quantum (NISQ) devices. Gate-based quantum devices require error correction on many physical qubits to form a logical qubit [11]. All quantum algorithms with exponential speedups require logical qubits.

NISQ annealers cannot guarantee global minima but can still provide low-energy solutions to the Ising Hamiltonian with a quantum advantage over classical simulated annealing [12]. NISQ annealers can be used as an efficient minimization tool by mapping problems to the Hamiltonian of the annealer. An iterative quantum annealing approach for finite element problems has been provided by [13]. Although quantum annealing can provide a quantum advantage over simulated annealing, exponential speedups are theorized on gate-based quantum computing architectures. In this paper, we focus on gate-based quantum computing.

Amplitude encoding is commonly used in gate-based quantum computing algorithms to represent data. This is achieved by encoding the entries of a complex vector $\boldsymbol{v}$ into the amplitudes of the basis vectors of a quantum state $|v\rangle$, which is normalized according to the Born rule $|\langle v|v\rangle|_2^2 = 1$. Basis encoding is an alternative that encodes a binary string of data as the correspondingly labeled basis state of a system of qubits.

A quantum linear system algorithm (QLSA) takes an amplitude-encoded quantum state $|\boldsymbol{b}\rangle$ as the input and outputs a state $|x\rangle$ proportional to the solution of a $d$-sparse system of linear equations $\boldsymbol{Ax} = \boldsymbol{b}$, where $\boldsymbol{A} \in \mathbb{C}^{N \times N}$, $\boldsymbol{x}, \boldsymbol{b} \in \mathbb{C}^N$, and $d$ denotes the maximum number of non-zero entries in any row or column. The first QLSA was developed by [1] with complexity $O(\log(N)\, d^2 \kappa^2 / \epsilon)$. Various improved complexities have been reported since then with a best known complexity of $O(d\kappa \operatorname{poly} \log(d\kappa N / \epsilon))$ using a sparse matrix query model and linear combination of unitary (LCU) techniques [14], which can be improved to $O(\kappa \operatorname{poly} \log(\kappa N/\epsilon))$ using a block-encoding query model combined with quantum signal processing (QSP). Furthermore, the algorithms can deal with singular systems of equations if $\boldsymbol{b}$ lies in the null space of $\boldsymbol{A}$. Reading out the complete quantum state $|\boldsymbol{x}\rangle$ requires $O(N)$ operations, which eliminates any exponential speedup w.r.t $N$ [1]. However, scalar properties of the quantum state may be extracted efficiently [3]. Identification of properties that can be efficiently extracted and algorithms for extraction are open problems [15], [16]. In this paper we focus on the problem of efficiently producing a quantum state $|\boldsymbol{x}\rangle$.

LCU and QSP QLSAs approximate $\boldsymbol{A}^{-1}$, where $\boldsymbol{A}$ is normalized s.t. $\|\boldsymbol{A}\| \leq 1$, with a Chebyshev polynomial approximation $P(\boldsymbol{A})$ over the interval $[-1, -1/\kappa] \cup [1/\kappa, 1]$ s.t. $\max \|P(\boldsymbol{A}) - \boldsymbol{A}^{-1}\| \leq \epsilon$. The LCU method applies the Chebyshev polynomial directly, leading to a larger qubit overhead due to 'select' and 'prepare' operations. In contrast, the QSP method uses the quantum signal processing theorem [2] to apply the polynomial using a QSP circuit, which requires the QSP sequence of rotation angles equivalent to the desired Chebyshev polynomial [2]. The QSP rotation angle sequence may be obtained using the QSPPACK library [17]. The number of Chebyshev polynomial terms, and equivalently the number of QSP rotation angles, scales $\propto \kappa \log 1/\epsilon$. Consequently, the quantum circuit depth grows linearly with $\kappa$. Furthermore, obtaining the QSP angles becomes increasingly difficult due to loss of numerical precision from increasingly higher order Chebyshev polynomial terms. Recent work has shown that symmetric positive-definite systems can be solved with $O(\sqrt{\kappa} \operatorname{poly} \log(\kappa N/\epsilon))$ scaling by approximating over the interval $[1/\kappa, 1]$ instead [18], which matches classical optimal scaling.

Several algorithms have been proposed to solve sparse systems of ordinary differential equations on quantum computers. The algorithms can be broadly classified by their application to linear or non-linear systems, and homogeneous or inhomogeneous systems. The first quantum algorithm for ordinary differential equations using the Euler method was proposed by [19] for nonlinear inhomogeneous systems by using $(16/\epsilon^2)^{\frac{t}{\Delta t}}$ copies of the initial condition, leading to a polynomial complexity with $\log(N)$ and an exponential complexity with $t$ and $\Delta t$. [20] proposed using higher-order methods for linear ordinary differential equations by constructing a block-encoded linear system using a Feynman's clock encoding and solving it using the HHL algorithm [1] to achieve a quadratic complexity in time. This was later improved using the LCU linear system algorithm and a Taylor series approximation of the 1st-order linear ordinary differential equation $\dot{\boldsymbol{x}} = \boldsymbol{A}\boldsymbol{x} + \boldsymbol{b}$ to obtain a complexity of $O(\kappa_V dT\|\boldsymbol{A}\|_2 \operatorname{poly}(\log(\kappa_V dg\beta T\|\boldsymbol{A}\|_2/\epsilon)))$ where $\kappa_V$ is the condition number of the eigenvalues, $d$ is the maximum number of non-zeros in any row or column, $T$ is total time, $g = \max_{t \in [0,T]} \||\boldsymbol{x}(t)\rangle\|/\||\boldsymbol{x}(T)\rangle\|$, and $\beta = (\||\boldsymbol{x}(0)\rangle\| + T\||\boldsymbol{b}\rangle\|)/\||\boldsymbol{x}(T)\rangle\|$. Subsequently [21] improved the number of copies required for nonlinear problems to quadratic and [22] used a Carleman linearization approach. A quantum machine learning approach was proposed by [23] using differentiable quantum circuits. Algorithms for homogeneous ordinary differential equations exhibit more favorable scaling [24], but their application to solving finite element problems is limited.

Extensions to partial differential equations were made by [25] and [26]. [25] solves the Laplace and second-order elliptic problems on square or rectangular domains with regularly spaced grid points using finite difference or spectral methods. [26] approaches the problem by inverting polynomial differential operators. A preconditioned algorithm for the finite element method was proposed by [15], where an SPAI (Sparse Approximate Inverse) left-preconditioner is used for improved scaling with respect to $\kappa$.

However, an explicit algorithm to form the preconditioner on quantum hardware is not provided. [3] point out in their analysis that while producing a quantum state proportional to the solution of partial differential equations using the finite element method can be exponentially efficient, extracting information about the output can eliminate the exponential speedups, but polynomial speedups are possible for problems with large higher-order derivatives and spatial dimensions.

Iterative solution methods have demonstrated great success in solving problems in mechanics. While the solution of a linear system using direct methods is $O(N^3)$ [27], optimal iterative methods like the multigrid method can scale as $O(N)$ [28]. However, iterative methods have not been explored in the context of quantum computing. A limiting factor is calculating inner products since it entails measuring a quantum state. However, relaxation or smoothing methods for positive-definite systems do not require calculating inner products [29], making them a viable option.

In this paper we present qRLS, the first iterative approach for the solution of semi-positive-definite systems by block-encoding a linear stationary iteration or relaxation method in a larger system. We show that the condition number of this large system is dependent only on the number of iterations and not the condition number or size of $A$. On classical architectures, this approach would be prohibitively expensive due to the large size of the block-encoded system. However, since the state space of qubits is exponentially large, the larger system can be accommodated with a few additional qubits on gate-based quantum computing architectures.

The relaxation techniques presented are strictly convergent and can be exponentially efficient on quantum computers with the number of degrees of freedom, requiring $O(\log(N))$ qubits, and $O(l \operatorname{poly} \log(l/\epsilon))$ or $O(dl \operatorname{poly} \log(dl/\epsilon))$ circuit depth (time complexity) for the block-encoded and sparse access models respectively, where $N$ is the number of degrees of freedom, $l$ is the number of relaxation steps, $d$ is the maximum number of nonzero entries in any row or column, and $\epsilon$ is the desired precision when using a QSP QLSA. Furthermore, the technique works for the general case where $b$ lies in the null space of $A$ since the null space is invariant to the relaxation technique.

The paper is organized as follows: Section 2 introduces the solution of finite element problems using iterative schemes used in our method. Section 3 presents the implementation of qRLS on gate-based quantum computers with an analysis of the steps in the algorithm and overall complexity. Section 4 provides numerical results from quantum computer simulators compared with results from classical solvers.

## 2. Finite element problem

We consider the following linear finite element problem defined on the open bounded domain $\Omega_h \in \mathbb{R}^d, d \in \{1,2,3\}$ with boundary $\Gamma_h$

*Find $x \in \mathbb{R}^N$ such that*

$$\boldsymbol{\Psi}(\boldsymbol{u}) = \mathbf{A}\boldsymbol{x} - \mathbf{b} = \mathbf{0} \tag{1}$$

subject to $x = x_g$ on $\Gamma_g$

where $x$ is a vector of nodal unknowns, $A \in \mathbb{R}^{N \times N}$ is the system matrix, $b \in \mathbb{R}^N$ is the forcing function, $N$ is the number of degrees of freedom of the discretized problem and $\Gamma_g$ is the Dirichlet boundary.

## 2.1. Iterative method

A general first order iterative solution method may be defined as

$$x^{(l+1)} = (I - \tau_l C^{-1} A) x^{(l)} + \tau_l C^{-1} b \qquad (2)$$

where $C \in \mathbb{R}^{N \times N}$ is a left preconditioning matrix and $\tau_l \in \mathbb{R}$ are constants for determined by the iteration scheme. Equation (2) is reduced to the Richardson Iteration for a choice of $C^{-1} = I$ and $\tau_l = \omega$ or the Jacobi iteration for a choice of $C^{-1} = diag(A)^{-1}$ and $\tau_l = \omega$ where $\omega$ is the damping factor. If $\tau_l = \tau$ is constant, the iterative method is stationary.

**Theorem 1** [29]: The sequence of vectors $x^{(l)}$ converges to the solution of $Ax = b$ for any $x^{(0)}$ if and only if $\rho(R) < 1$ where $R = I - \tau C^{-1} A$.

Given a positive-definite $C^{-1}A$ the optimal choice $\tau = \frac{2}{\lambda_{min} + \lambda_{max}}$ for a stationary method and the asymptotic convergence factor is $\rho(R) = \frac{1 - \lambda_{min}/\lambda_{max}}{1 + \lambda_{min}/\lambda_{max}}$ where $\lambda_{max}, \lambda_{min}$ are the maximum and minimum (non-zero) eigenvalues of $C^{-1}A$. When $R$ is symmetric and positive-semidefinite $\rho(R) = \|R\|$ is the average convergence factor, and the convergence is monotonic.

**Note**: $\|\cdot\|$ denotes the 2-norm throughout this paper unless stated otherwise.

For symmetric positive-definite $R$, the relative error $\|e_l\|/\|e_0\|$ is decreased monotonically by a factor of $\epsilon > 0$ in $l = \left\lceil \frac{1}{2} \frac{\lambda_{max}}{\lambda_{min}} \ln \frac{1}{\epsilon} \right\rceil$ iterations.

The convergence rate may be improved by an order of magnitude using the non-stationary 1st order Chebyshev iterative method to require $l = \left\lceil \frac{1}{2} \sqrt{\frac{\lambda_{max}}{\lambda_{min}}} \ln \frac{1}{\epsilon} \right\rceil$ iterations. The Chebyshev method achieves optimal convergence with the choice of $\tau_i^{-1}$ chosen to be the Chebyshev nodes in the domain $\lambda \in [a, b]$ and $a \leq \lambda_{min}$, $b \leq \lambda_{max}$. However, the 1st order Chebyshev iterative method can be numerically unstable.

We note that the iterative method is capable of handling semi-definite systems since the null space of $C^{-1}A$ is equivalent to $I$ for $I - \tau C^{-1} A$. For this paper, we consider stationary methods for systems for which bounds $a \leq \lambda_{min}, \lambda_{max} \leq b$ are known with the choice $C^{-1} = I$.

## 3. Stationary linear iterations on quantum computers

In this section, we present the steps of the qRLS algorithm for iterative solution of finite element problems. We introduce the standard bra-ket or Dirac notation for $\langle \psi |$ and $|\phi\rangle$ in Section 3.1.. The first step, described in Section 3.2, is to define a block encoded linear system problem $M_{l,c}|x\rangle = |y\rangle$ whose solution provides the sequence of iterates $|x\rangle$. Next, we prepare the quantum state $|y\rangle$, presented in Section 3.3. The linear system $M_{l,c}|x\rangle = |y\rangle$ is then solved using a QLSA to produce the quantum state $|x\rangle$ with a precision of $\delta$, which encodes the sequence of iterations as $|x\rangle = \sum_{i=0}^{l+c} |i\rangle |x^{(i)}\rangle$, as detailed in Section 3.4. To produce the quantum state of the desired final iterate $|x^{(i)}\rangle$ where $i \in [l, l+c]$ a measurement is performed on the first register $|i\rangle$, which will yield a basis state $i \in [l, l+c]$ with a probability $p$ which is analyzed in Section 3.5.

In Section 3.6 we consider the effects of the QLSA error $\delta$ on the precision of the final output after measuring the first register and the success probability. Since quantum states are always normalized, to obtain the quantum state $|x^{(i)}\rangle$ with precision $\epsilon$ the solution $|x\rangle$ from a QLSA must be obtained with a higher precision $\delta \leq \epsilon$. Furthermore, the error $\delta$ in the QLSA output affects the success probability, which is accounted for to get an updated success probability $p'$ for the QLSA output.

Finally, we arrive at the overall complexity of the qRLS algorithm in Section 3.7 by individually considering the complexity of solving the linear system $M_{l,c}|x\rangle = |y\rangle$, the success probability $p'$, the required precision $\delta$, and the complexity of preparing the state $|y\rangle$.

### 3.1. Notation

A quantum state of qubits is described as a 'bra' vector $|\psi\rangle$, where $\psi$ may be replaced with a descriptor of the state e.g. an integer $i$ to denote the standard basis state $|i\rangle$ corresponding to a vector $e_i$, or a vector $x$ to denote a quantum state encoding $|x\rangle = x/\|x\|$ as the probability amplitudes of $|x\rangle$. A 'bra' is the conjugate transpose of a ket and is denoted as $\langle x| = ((|x\rangle)^T)^*$. $|\psi\rangle|\phi\rangle$ is a Kronecker product $|\psi\rangle \otimes |\phi\rangle$.

A quantum state $|\chi\rangle$ is always normalized as $|\langle\chi|\chi\rangle|^2 = 1$ and can be written as $|\chi\rangle = \frac{1}{\sqrt{|a|^2+|b|^2}}(\alpha|\psi_a\rangle|\psi_b\rangle + \beta|\phi_a\rangle|\phi_b\rangle)$. For brevity and convenience, we occasionally omit the overall normalization constants $\frac{1}{\sqrt{|a|^2+|b|^2}}$ of quantum states without any loss of generality.

### 3.2. Block encoding of matrix iterations.

We encode the iterative processes as a back substitution in a larger block bidiagonal linear system. $l$ is the number of iteration steps and $c$ is the number of copies of the final iterate $x^{(l)}$. The choice of $l$ and $c$ affects the condition number of the linear system and determines the success probability of the algorithm, which are discussed in Sections 3.3 and 3.4.

For the 1st order iterative method, we define the matrix $M_{l,c} \in \mathbb{R}^{(l+c+1)N \times (l+c+1)N}$:

$$M_{l,c} = \sum_{i=0}^{l+c} |i\rangle\langle i| \otimes I - \sum_{i=1}^{l} |i\rangle\langle i-1| \otimes R - \sum_{i=l+1}^{l+c} |i\rangle\langle i-1| \otimes I \tag{3}$$

where $R = I - \tau A$. $\tau$ is chosen to ensure $\|\tau A\| \leq 1$ and hence (due to positive-definiteness) $\|R\| \leq 1$ to solve the linear system.

$$M_{l,c}|x\rangle = |y\rangle \tag{4}$$

where $|y\rangle = \sum_{i=1}^{l} |i\rangle\tau|b\rangle + |0\rangle|x_0\rangle$, and $|x\rangle = \sum_{i=0}^{l+c} |i\rangle|x_i\rangle$ where $|x_i\rangle = x^{(i)}$, the $i^{th}$ iterate, for $i \in [0, l]$ or a copy of the final iterate $x^{(l)}$ for $i \in [l+1, l+c]$. We refer to the block-indexing register of qubits $|i\rangle$ as the 'first register' and the remaining qubits as the 'second register'.

As an example, $M_{2,2}|x\rangle = |y\rangle$ forms the system

$$\begin{bmatrix} I & & & & \\ -R & I & & & \\ & -R & I & & \\ & & -I & I & \\ & & & -I & I \end{bmatrix} \begin{bmatrix} x^{(0)} \\ x^{(1)} \\ x^{(2)} \\ x^{(3)} \\ x^{(4)} \end{bmatrix} = \begin{bmatrix} x^{(0)} \\ \tau b \\ \tau b \\ 0 \\ 0 \end{bmatrix}. \tag{5}$$

### 3.3. State preparation

Since we are interested in solving the linear system $M_{l,c}|x\rangle = |y\rangle$, we need to be able to prepare the state $|y\rangle = \sum_{i=1}^{l} |i\rangle \tau |b\rangle + |0\rangle |x_{in}\rangle$ efficiently. A quantum oracle is a black-box method to access information. $|y\rangle$ can be prepared using oracles $\mathcal{O}_x$ and $\mathcal{O}_b$ for $|x_{in}\rangle$ and $|b\rangle$ respectively. The oracles $\mathcal{O}_x$ and $\mathcal{O}_b$ are unitary operations that perform the map $\mathcal{O}_x|1\rangle|\psi\rangle = |1\rangle|\psi\rangle$, $\mathcal{O}_x|0\rangle|0\rangle = |0\rangle|x_{in}\rangle$ and $\mathcal{O}_x|0\rangle|\psi\rangle = |0\rangle|\psi\rangle$, $\mathcal{O}_b|1\rangle|0\rangle = |1\rangle|b\rangle$. Given access to efficient implementations of these oracles, the state $|y\rangle = \sum_{i=1}^{l} |i\rangle \tau |b\rangle + |0\rangle |x_{in}\rangle$ can be prepared efficiently.

**Lemma 2:** [30] Let $\mathcal{O}_x$ be a unitary that maps $|1\rangle|\psi\rangle$ to $|1\rangle|\psi\rangle$ for any $|\psi\rangle$ and maps $|0\rangle|0\rangle$ to $|0\rangle|\bar{x}_{in}\rangle$ where $\bar{x}_{in} = x_{in}/\|x_{in}\|$. Let $\mathcal{O}_b$ be a unitary that maps $|0\rangle|\psi\rangle$ to $|0\rangle|\psi\rangle$ for any $|\psi\rangle$ and maps $|1\rangle|0\rangle$ to $|1\rangle|\bar{b}\rangle$ where $\bar{b} = b/\|b\|$. Suppose we know $\|x_{in}\|$ and $\|b\|$. Then the state proportional to $\sum_{i=1}^{l} |i\rangle \tau |b\rangle + |0\rangle|x_{in}\rangle$ can be produced with $O(1)$ calls to $\mathcal{O}_x$ and $\mathcal{O}_b$, and $O(\text{poly} \log l)$ elementary quantum gates.

We use Lemma 2 to factor in the complexity of preparing the state $|y\rangle$. In Algorithm 1, for the restart step we do not use $\mathcal{O}_x$ and instead skip to the next step to prepare $|1\rangle|b\rangle$ using $\mathcal{O}_b$ and follow the subsequent steps.

### 3.4. Linear system solution

The matrices $M_{l,c}$ are block bidiagonal. Furthermore, we see that $\|R\| \leq 1$ for the sub-diagonal blocks for symmetric positive-definite $A$. In Lemma 3, we provide bounds on the norm and condition number of $M_{l,c}$. In Lemma 4, we provide the complexity of solving the QLSA using the QSP algorithm.

**Lemma 3:** $\forall M \in \mathbb{R}^{lN \times lN}$ s.t. $M_{ik} \in \mathbb{R}^{N \times N}$ where $M_{ii} = I, M_{i+1,i} = -A_i, M_{ji} = 0$ and $\forall i, j, k \in [1, l] \subset \mathbb{N}$ where $j \neq i, i+1$ and $\|A_i\| \leq 1 \, \forall i$

$$\|M\| \leq 2, \|M^{-1}\| \leq l \tag{6}$$

$$\kappa_M \leq 2l \tag{7}$$

**Proof**:

$$M = \begin{bmatrix} I & & & & \\ -A_1 & I & & & \\ & -A_2 & I & & \\ & & \ddots & \ddots & \\ & & & -A_{l-1} & I \end{bmatrix} \tag{8}$$

$$M^{-1} = \begin{bmatrix} I & & & & & \\ A_1 & I & & & & \\ A_2 A_1 & A_2 & I & & & \\ & & & \ddots & \ddots & \\ \prod_{i=l-1}^{1} A_i & \cdots & & A_{l-2} A_{l-1} & A_{l-1} & I \end{bmatrix} \tag{9}$$

Denoting $[S]_{N,k} \in \mathbb{R}^{N \times N}$ as the matrix formed by extracting the $k^{th}$ $N \times N$ block sub diagonal of $M$

$$\left\| \prod_{i=j}^{k} A_i \right\| \leq 1 \forall j, k \because \|A_i\| \leq 1 \ \forall \ i \in [1, l-1] \tag{10}$$

$$\|M\| = \left\| [S]_{N,0} + [S]_{N,1} \right\| \leq \left\| [S]_{N,0} \right\| + \left\| [S]_{N,1} \right\| \leq 2 \tag{11}$$

$$\|M^{-1}\| = \left\| \sum_{k=0}^{l} [S]_{N,l} \right\| \leq \sum_{k=0}^{l} \left\| [S]_{N,l} \right\| \leq l \tag{12}$$

$$\kappa_M = \|M\| \|M^{-1}\| \leq 2l \quad \square \tag{13}$$

We also note that $M$ is positive-definite since the eigenvalues of a triangular matrix lie on its diagonal, and $diag(M) = I$. Furthermore, we note that the Hermitian dilation of $M$, $\bar{M} = \begin{bmatrix} 0 & M \\ M^T & 0 \end{bmatrix}$ is positive-definite and symmetric, with eigenvalues being the singular values of $M$. $A_i$, $M$, and $\bar{M}$ have the same $O(d)$ sparsity where $d$ is the maximum number of non-zero entries in any row or column. The equivalent system $\begin{bmatrix} 0 & M \\ M^T & 0 \end{bmatrix} \begin{bmatrix} 0 \\ x \end{bmatrix} = \begin{bmatrix} y \\ 0 \end{bmatrix}$ can be solved to obtain $\begin{bmatrix} 0 \\ x \end{bmatrix} = \begin{bmatrix} 0 \\ M^{-1} y \end{bmatrix}$ Finally, we note that from Lemma 3, we have bounds on the singular values of $M$ as $\sigma_{max} \leq 2$ and $\sigma_{min}^{-1} \geq l$.

Using these facts, we now establish Lemma 4 for the complexity of solving the linear system $M_{l,c}|x\rangle = |y\rangle$.

### 3.5. Success probability

Solving the linear system $M_{l,c}|x\rangle = |y\rangle$ produces a quantum state encoding the sequence of iterations $|x\rangle = \sum_{i=0}^{l+c} |i\rangle |x^{(i)}\rangle$. The desired output, $|x^{(i)}\rangle \ \forall \ i \in [l, l+c]$, can be extracted by performing a measurement on the first register $|i\rangle$. Measuring the first register $|i\rangle$ in a basis state $j$ will indicate that the second register $|x^{(i)}\rangle$ is now in the state $|x^{(j)}\rangle$. This may be expressed mathematically as:

$$(\langle j | \otimes \mathbb{I}) \left( \sum_{i=0}^{l+c} |i\rangle |x^{(i)}\rangle \right) = \frac{1}{\sqrt{1 - \sum_{\substack{i=0 \\ \neq j}}^{l+c} \||x^{(i)}\rangle\|^2}} |x^{(j)}\rangle \tag{14}$$

By definition of a quantum state, the probability $p_j$ of measuring the first register in the state $|j\rangle$ is $p_j = \||x^{(j)}\rangle\|^2$. Since we are interested in producing the state $|x^{(i)}\rangle$ where $i \in [l, l+c]$, we provide in the following lemmas the probability of successfully measuring the first register $|i\rangle$ in a basis state $i \in [l, l+c]$.

The state $|x\rangle = \sum_{i=0}^{l+c} |i\rangle |x^{(i)}\rangle$ can be represented as $|x\rangle = |x_{good}\rangle + |x_{bad}\rangle$ where $|x_{good}\rangle = \sum_{i=l}^{l+c} |i\rangle |x^{(i)}\rangle$ is the desired output and copies. Using the Born rule for quantum states:

$$\||x\rangle\|^2 = \||x_{good}\rangle\|^2 + \||x_{bad}\rangle\|^2 = 1 \tag{15}$$

In Lemmas 5 and 6 we provide bounds on the success probability of obtaining the desired output $|x^{(i)}\rangle \ \forall \ i \in [l, l+c]$ when a measurement is performed on the first register for the case $x_{in} = \mathbf{0}$ and $x_{in} \neq \mathbf{0}$, respectively.

**Lemma 4**: Given a monotonically convergent scheme and starting with an initial guess of $\mathbf{0}$ with a choice of $c = l - 1$, successfully measuring the first register $|i\rangle$ in the state such that $i \in [l, l+c]$ has probability $p \geq 1/2$.

**Proof**:

Starting with an initial guess of $\mathbf{0}$, the solution grows monotonically if the error decreases monotonically. This is without any loss of generality since instead of solving $Ax = b$ the equivalent system $Ae = r$ can be solved for the error where $r = f - Ax$. Hence

$$\frac{\||x_{good}\rangle\|^2}{c+1} \geq \frac{\||x_{bad}\rangle\|^2}{l} \tag{16}$$

$$p = \frac{\||x_{good}\rangle\|^2}{\||x\rangle\|^2} = \frac{\||x_{good}\rangle\|^2}{\||x_{good}\rangle\|^2 + \||x_{bad}\rangle\|^2} \geq \frac{c+1}{l+c+1} \tag{17}$$

with $c = l - 1$ \hfill (18)

$$p \geq \frac{1}{2} \qquad \square \tag{19}$$

**Lemma 5**: Given a monotonically convergent scheme starting with an initial guess $x_{in}$ and error $\epsilon = \|\tilde{x} - x_{in}\| \leq \|\tilde{x}\|$, where $\tilde{x} = A^{-1}b$, and a choice of $c = l - 1$, the probability $p$ of successfully measuring the first register $|i\rangle$ in the state such that $i \in [l, l+c]$ is $p \geq \frac{1}{2}\left(\frac{\|\tilde{x}\| - \epsilon}{\|\tilde{x}\| + \epsilon}\right)^2$.

**Proof**:

Since the scheme is monotonically convergent, $\epsilon \leq \|\tilde{x} - x_i\| \leq \|\tilde{x} - x_{in}\| \ \forall \ i \in [0, l+c]$.

$$\|\tilde{x}\| - \epsilon \leq \|x_i\| \leq \|\tilde{x}\| + \epsilon \ \forall \ i \text{ when } \epsilon \leq \|\tilde{x}\| \tag{20}$$

Using the choice $c = l - 1$

$$(\|\tilde{x}\| - \epsilon)^2 \leq \frac{\||x_{good}\rangle\|^2}{l} \leq (\|\tilde{x}\| + \epsilon)^2 \tag{21}$$

$$\frac{\||x_{bad}\rangle\|^2}{l} \leq (\|\tilde{x}\| + \epsilon)^2 \tag{22}$$

$$\frac{l}{\||x_{good}\rangle\|^2 + \||x_{bad}\rangle\|^2} \geq \frac{1}{2(\|\tilde{x}\| + \epsilon)^2} \tag{23}$$

Multiplying (21) by (23)

$$p = \frac{\||x_{good}\rangle\|^2}{\||x_{good}\rangle\|^2 + \||x_{bad}\rangle\|^2} \geq \frac{1}{2}\left(\frac{1-\epsilon/\|\tilde{x}\|}{1+\epsilon/\|\tilde{x}\|}\right)^2 \qquad \square \qquad (24)$$

### 3.6. Precision

To produce the state $|x_l\rangle = |x^{(i)}\rangle \ \forall i \in [l, l+c]$ with precision $\epsilon$, the QLSA must produce $|x\rangle = \sum_{i=0}^{l+c}|i\rangle|x^{(i)}\rangle$ with a higher precision $\delta \leq \epsilon$ due to normalization of quantum states after measurement of the first register as seen in Equation 14. The bounds on the required precision $\delta$ are presented in Lemmas 9 and 10 for $x_{in} = 0$ and $x_{in} \neq 0$, respectively. The intermediate Lemmas 7 and 8 from [30] are used to place a bound on the error $\epsilon$ of the renormalized quantum state after measurement and the success probability $p'$ respectively in the presence of error $\delta$.

We use the notation $|\overline{\phi}\rangle = \frac{|\phi\rangle}{\||\phi\rangle\|}$ to denote normalization.

Lemma 7 allows us to use a bound on the norm $\alpha$ of the exact desired iterate and a bound on the error of the QLSA output $\||x'\rangle - |x\rangle\| \leq \delta$ to bound the error $\||\overline{x}_l\rangle - |\overline{x}'_l\rangle\|$ of the actual iterate after measurement.

The error $\delta$ also affects the norm of the actual desired iterate $\alpha'$. Given bounds on $\delta$ and $\alpha$, Lemma 8 allows us to bound $\alpha'$ for an updated success probability in the presence of error $\delta$.

**Lemma 6**:[30] Let $|\psi\rangle = \alpha|0\rangle|\psi_0\rangle + \sqrt{1-\alpha^2}|1\rangle|\psi_1\rangle$ and $|\phi\rangle = \beta|0\rangle|\phi_0\rangle + \sqrt{1-\beta^2}|1\rangle|\phi_1\rangle$, where $|\psi_0\rangle, |\psi_1\rangle, |\phi_0\rangle$, and $|\phi_1\rangle$ are unit vectors, and $\alpha, \beta \in [0,1]$. Suppose $\||\psi\rangle - |\phi\rangle\| \leq \delta < \alpha$. Then

$$\||\phi_0\rangle - |\psi_0\rangle\| \leq \frac{2\delta}{\alpha-\delta} \qquad (25)$$

**Lemma 7**:[30] Let $|\psi\rangle = \alpha|0\rangle|\psi_0\rangle + \sqrt{1-\alpha^2}|1\rangle|\psi_1\rangle$ and $|\phi\rangle = \beta|0\rangle|\phi_0\rangle + \sqrt{1-\beta^2}|1\rangle|\phi_1\rangle$, where $|\psi_0\rangle, |\psi_1\rangle, |\phi_0\rangle$, and $|\phi_1\rangle$ are unit vectors, and $\alpha, \beta \in [0,1]$. Suppose $\||\psi\rangle - |\phi\rangle\| \leq \delta < \alpha$. Then

$$\beta \geq \alpha - \delta \qquad (26)$$

We now present Lemmas 9 and 10 to obtain bounds on the required precision $\delta$ for a final precision $\epsilon$ after measurement of the first register and the success probability in the presence of error $\delta$, for the cases $x_{in} = 0$ and $x_{in} \neq 0$ respectively.

**Lemma 8**: Given $x_{in} = 0$, the QLSA outputs the state $|x'\rangle$ with precision $\delta$. Using a choice of $\delta \leq \frac{\epsilon}{2\sqrt{2l}}$ the state $|x'_l\rangle$ can be obtained with precision $\epsilon$ with a probability of success $p' \geq \frac{1-\epsilon}{2}$

**Proof**:

Defining $|x\rangle = \sum_{i=0}^{l+c} \alpha_i |i\rangle|\overline{x}_i\rangle$

from Lemma 5, $\forall i \in [l, l+c]$ the success probability $p$ of measuring the first register in any of the states $|i\rangle$ is $p \geq \frac{1}{2}$ for a choice of $c = l - 1$. We define $p_i$ as the probability of measuring a particular state $|i\rangle$ for $i \geq l$.

By definition of the probability amplitudes of a quantum state $p_i = (\alpha_i)^2 \geq \frac{1}{2l}$ for $l \leq i \leq 2l - 1$ for a choice of $c = l - 1$

Suppose a QLSA outputs $|\bar{x}'\rangle$ s.t.

$$\||\bar{x}\rangle - |\bar{x}'\rangle\| \leq \delta < \alpha_l \leq 1 \tag{27}$$

Using Lemmas 7 and 8

$$\||\bar{x}_l\rangle - |\bar{x}'_l\rangle\| \leq \frac{2\delta}{\alpha_l - \delta} \leq \frac{2\delta}{\frac{1}{\sqrt{2l}} - \delta} \leq \epsilon < 1 \text{ where } \epsilon > 0 \tag{28}$$

where $\epsilon$ is the desired precision for $|\bar{x}'_l\rangle$

$$(2 + \epsilon)\delta \leq \frac{\epsilon}{\sqrt{2l}} \tag{29}$$

$$2\delta \leq \frac{\epsilon}{\sqrt{2l}} \tag{30}$$

$$0 < \delta \leq \frac{\epsilon}{2\sqrt{2l}} < 1 \tag{31}$$

and $\alpha'_l \geq \alpha_l - \delta \geq \frac{1}{\sqrt{2l}} - \delta$ \hfill (32)

Therefore, the success probability of measuring the 1st register in the state $|l'\rangle$ where $l' \in [l, 2l - 1]$ is

$$p' = l\alpha'^2_l \geq l\left(\frac{1}{\sqrt{2l}} - \delta\right)^2 = \frac{1}{2} + l\delta^2 - \sqrt{2l}\delta \geq \frac{1}{2} - \sqrt{2l}\delta. \tag{33}$$

By choosing $\delta \leq \frac{\epsilon}{2\sqrt{2l}}$ \hfill (34)

$$p' \geq \frac{1-\epsilon}{2}. \qquad \square \tag{35}$$

**Lemma 9**: Given $x_{in}$ s.t. $\|x_{in} - \tilde{x}\| \leq \epsilon_1$, the QLSA outputs the state $|x'\rangle$ with precision $\delta$. Using a choice of $\delta \leq \frac{\epsilon_2}{2\sqrt{2l}}\left(\frac{1-\epsilon_1/\|\tilde{x}\|}{1+\epsilon_1/\|\tilde{x}\|}\right)$, the state $|x'_l\rangle$ can be obtained with precision $\epsilon_2$ with a probability of success

$$p' \geq \frac{1-\epsilon_2}{2}\left(\frac{1-\frac{\epsilon_1}{\|\tilde{x}\|}}{1+\frac{\epsilon_1}{\|\tilde{x}\|}}\right)^2$$

**Proof**:

Defining $|x\rangle = \sum_{i=0}^{l+c} \alpha_i|i\rangle|\bar{x}_i\rangle$

from Lemma 6, $\alpha_i \geq \frac{1}{\sqrt{2l}}\left(\frac{1-\epsilon_1/\|\tilde{x}\|}{1+\epsilon_1/\|\tilde{x}\|}\right)$ for $l \leq i \leq 2l - 1$ for a choice of $c = l - 1$

Suppose a QLSA outputs $|\bar{x}'\rangle$ s.t.

$$\||\bar{x}\rangle - |\bar{x}'\rangle\| \leq \delta < \alpha_l \leq 1 \tag{36}$$

Using Lemmas 7 and 8

$$\||\bar{x}_l\rangle - |\bar{x}'_l\rangle\| \leq \frac{2\delta}{\alpha_l - \delta} \leq \frac{2\delta}{\frac{1}{\sqrt{2l}}\left(\frac{1-\frac{\epsilon_1}{\|\tilde{x}\|}}{1+\frac{\epsilon_1}{\|\tilde{x}\|}}\right) - \delta} \leq \epsilon_2 < \epsilon_1 \text{ where } \epsilon_2 > 0 \tag{37}$$

where $\epsilon_2$ is the desired precision for $|\bar{x}'_l\rangle$

$$(2+\epsilon_2)\delta \leq \frac{\epsilon_2}{\sqrt{2l}}\left(\frac{1-\epsilon_1/\|\tilde{x}\|}{1+\epsilon_1/\|\tilde{x}\|}\right) \tag{38}$$

$$2\delta \leq \frac{\epsilon_2}{\sqrt{2l}}\left(\frac{1-\epsilon_1/\|\tilde{x}\|}{1+\epsilon_1/\|\tilde{x}\|}\right) \tag{39}$$

$$0 < \delta \leq \frac{\epsilon_2}{2\sqrt{2l}}\left(\frac{1-\epsilon_1/\|\tilde{x}\|}{1+\epsilon_1/\|\tilde{x}\|}\right) < 1 \tag{40}$$

and $\alpha'_l \geq \alpha_l - \delta \geq \frac{1}{\sqrt{2l}}\left(\frac{1-\epsilon_1/\|\tilde{x}\|}{1+\epsilon_1/\|\tilde{x}\|}\right) - \delta \tag{41}$

Therefore, the success probability of measuring the 1st register in the state $|l'\rangle$ where $l' \in [l, 2l-1]$ is

$$p' = l\alpha'^2_l \geq l\left(\frac{1}{\sqrt{2l}}\left(\frac{1-\epsilon_1/\|\tilde{x}\|}{1+\epsilon_1/\|\tilde{x}\|}\right) - \delta\right)^2 = \frac{1}{2}\left(\frac{1-\frac{\epsilon_1}{\|\tilde{x}\|}}{1+\frac{\epsilon_1}{\|\tilde{x}\|}}\right)^2 + \frac{\delta^2}{2} - \sqrt{2l}\delta\left(\frac{1-\frac{\epsilon_1}{\|\tilde{x}\|}}{1+\frac{\epsilon_1}{\|\tilde{x}\|}}\right) \geq \frac{1}{2} - \sqrt{2l}\delta\left(\frac{1-\frac{\epsilon_1}{\|\tilde{x}\|}}{1+\frac{\epsilon_1}{\|\tilde{x}\|}}\right). \tag{42}$$

By choosing $\delta \leq \frac{\epsilon_2}{2\sqrt{2l}}\left(\frac{1-\epsilon_1/\|\tilde{x}\|}{1+\epsilon_1/\|\tilde{x}\|}\right) \tag{43}$

$$p' \geq \frac{1-\epsilon}{2}\left(\frac{1-\frac{\epsilon_1}{\|\tilde{x}\|}}{1+\frac{\epsilon_1}{\|\tilde{x}\|}}\right)^2. \qquad \square \tag{44}$$

### 3.7. Complexity analysis of the qRLS algorithm

The overall qRLS algorithm consists of preparing the state $|y\rangle$, solving the QLSP $M_{l,c}|x\rangle = |y\rangle$ using either the QSP QLSA or the PD-QLSA to a precision of $\delta$, and measuring the first register to obtain the desired iterate $|x_l\rangle$. If the first register $|i\rangle$ is measured in the required state $i \in [l, 2l-1]$, the second register is in the output state $|x_l\rangle$ of the iterative process after $l$ iterations. Otherwise, the second register contains the output $|x_i\rangle$ of the iterative process after $i < l$ iterations, and the process may be restarted with $\mathbf{0}$ or $|x_i\rangle$ as the initial guess.

---
Algorithm 1. Pseudo-code of qRLS

**Given** $\mathcal{O}_x, \mathcal{O}_b$ and $\mathcal{O}_M$ or $\mathcal{O}_P$
**Initialize**: success=false
**while** not(success) do
   Prepare state $|y\rangle$
   Solve $M_{l,l-1}|x\rangle = |y\rangle$ using QSP QLSA
   Measure the first register $|i\rangle$
   **if** $i \in [l, 2l-1]$
     success = True
   **else**
     Discard $|x_i\rangle$ or Restart with $|x_{in}\rangle = |x_i\rangle$
**end while**
---

For our analysis we consider the case where $|x_i\rangle$ is discarded if the measurement of the index register is unsuccessful, i.e. $i \notin [l, 2l-1]$.

**Theorem 10:** (Main result) Given an initial guess of $\mathbf{0}$, oracles $\mathcal{O}_x, \mathcal{O}_b$ to prepare the states $|x_{in}\rangle$ and $|b\rangle$, and an oracle $\mathcal{O}_M$ for the matrix $M_{l,l-1}$, the output $|x^{(l)}\rangle$ corresponding to $l$ linear stationary iterations for a symmetric positive-semidefinite matrix system $A$ can be prepared with precision $\epsilon \leq \frac{1}{2}$ with an overall complexity $O\left(l \operatorname{poly} \log\left(\frac{lN}{\epsilon}\right)\right)$.

**Proof**:

Using Lemma 2, the state $|y\rangle = \sum_{i=1}^{l} |i\rangle \tau |b\rangle + |0\rangle |x_{in}\rangle$ can be prepared in $O(\operatorname{poly} \log l)$ time.

We now solve the QLSP $M_{l,l-1}|x\rangle = |y\rangle$ where $|x\rangle = \sum_{i=0}^{2l-1} |i\rangle |x_i\rangle$ and $|x_i\rangle \propto x^{(i)}$, the $i^{th}$ iterate, for $i \in [0, l]$ or a copy of the final iterate $x^{(l)}$ for $i \in [l+1, 2l-1]$.

From Lemma 4, given an oracle $\mathcal{O}_M$ that block-encodes $M_{l,l-1}$ this QLSP can be solved using the QSP QLSA to a precision of $\delta$ with an overall complexity of $(l \operatorname{poly} \log(lN/\delta))$ to get $|x'\rangle$ where $\||x'\rangle - |x\rangle\| \leq \delta$.

To prepare the desired quantum state $|x'_l\rangle$ with error $\||x'_l\rangle - |x_l\rangle\| \leq \epsilon$ corresponding to the final iterate $x_l$, we perform a measurement on the first register. From Lemma 9, to satisfy $\||x'_l\rangle - |x_l\rangle\| \leq \epsilon$ the probability $p$ of measuring the first register $|i\rangle$ in the desired state $i \in [l, 2l-1]$ is $p' \geq \frac{1-\epsilon}{2}$. Also, to satisfy $\||x'_l\rangle - |x_l\rangle\| \leq \epsilon$, from Lemma 9 $\delta \leq \frac{\epsilon}{2\sqrt{2l}}$.

$\because \epsilon \leq \frac{1}{2},\ p' = O(1)$

We now arrive at individual complexities of $c_1 = O(\operatorname{poly} \log l)$ for state preparation, $c_2 = O(l \operatorname{poly} \log(lN/\epsilon))$ for linear system solution, and $c_3 = O(1)$ for probability of success.

The overall complexity is obtained by combining these to obtain $\frac{c_1 c_2}{c_3} = O(l \operatorname{poly} \log(lN/\epsilon))$. □

## 4. Numerical Examples

We first present results for problems in one- and two- dimensions to demonstrate the convergence of our block-encoded method using the QSP QLSA implemented in the Qiskit quantum simulation library [31] by comparing with classical iterative solutions.

The QSP QLSA requires calculation of phase angles to approximate $P(M) \approx M^{-1}$, given $\kappa$ and $\delta$. We use the QSPPACK library [17] to produce phase angles for $\kappa \leq 50$ and $\delta \leq \|P(A) - A^{-1}\| \leq 10^{-9}$ over the interval $\left[-1, -\frac{1}{\kappa}\right] \cup \left[\frac{1}{\kappa}, 1\right]$. The phase angles are problem-independent and can be reused for any problem with $\kappa \leq 50$ to obtain a precision of $\delta \leq 10^{-9}$. Since the matrix $M$ is not Hermitian, the Hermitian dilation $\begin{bmatrix} 0 & M \\ M^T & 0 \end{bmatrix} \begin{bmatrix} 0 \\ x \end{bmatrix} = \begin{bmatrix} y \\ 0 \end{bmatrix}$ is solved instead, which has the same condition number. Due to memory constraints, we restrict ourselves to smaller systems. This issue is not expected on a fault tolerant quantum system where the memory requirements scale as $\log N$ qubits.

We then present classical iterative methods to demonstrate the convergence behavior of linear stationary iterations on symmetric positive-semidefinite systems.

### 4.1. Poisson's problem in 1D

We consider the following Poisson's problems in 1D

$Find\ u: \overline{\Omega} \to \mathbb{R}\ such\ that$
$\frac{d^2u}{dx^2} - b = 0\ in\ \Omega \in (0, L)$  (45)
$with\ b = 1\ and\ u = g\ on\ \Gamma_g$

for the cases listed in Table 1.

*Table 1 Cases a and b considered for problems in 1D*

|  | $g, \Gamma_g$ |
|---|---|
| Case a | $u(0) = u(L) = 0$ |
| Case b | $u(0) = 0$ |

In Figure 1 we show the exact solutions of Cases (a) and (b) for $N = 1024$ nodes and a linear stationary iterative solution with an initial guess of $\mathbf{0}$ converged to a relative error $\frac{\|e\|}{\|e_0\|} \leq 10^{-8}$.

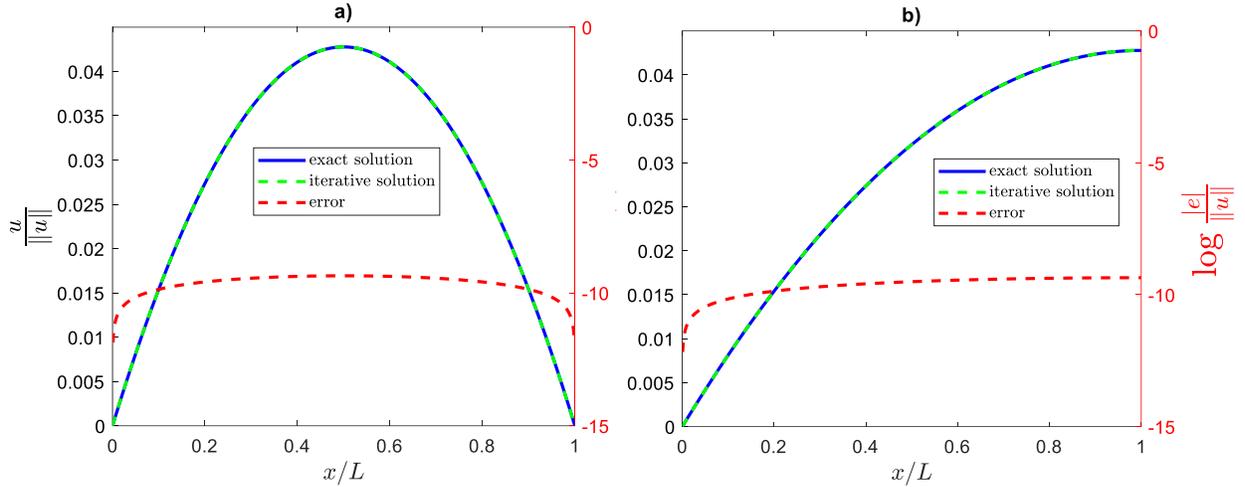

*Figure 1 Exact and converged iterative solutions of Cases (a) and (b).*

In Figure 2 we present results of our implementation of the QSP algorithm for QLSA using Qiskit's statevector simulator to obtain the quantum state encoding the iterative solution and compare with classical linear stationary iterations. Since quantum algorithms produce normalized vectors, for each $|x^{(l)}\rangle = \frac{x_l}{\|x_l\|}$ corresponding to the $l^{th}$ iterate $x_l$, we plot the convergence of the error $\frac{\|e_l\|}{\|e_0\|} = \frac{\left\|\frac{x_l}{\|x_l\|} - \frac{A^{-1}b}{\|A^{-1}b\|}\right\|}{\left\|0 - \frac{A^{-1}b}{\|A^{-1}b\|}\right\|}$. We also plot $\left\|\frac{x_{l_{quantum}}}{\|x_{l_{quantum}}\|} - \frac{x_{l_{classical}}}{\|x_{l_{classical}}\|}\right\|$, the error of a quantum iterate, where the classical iterate is the exact iterate. As expected for linear stationary iterations, the error monotonically decreases with every iteration. The error between the quantum and classical iterates stays below $10^{-9}$, as expected from the QSP QLSA based on our choice of $\delta \leq 10^{-9}$. We simulate the QSP QLSA for various choices of $l$ and $N$ that can satisfy the memory requirements of Qiskit's quantum simulator.

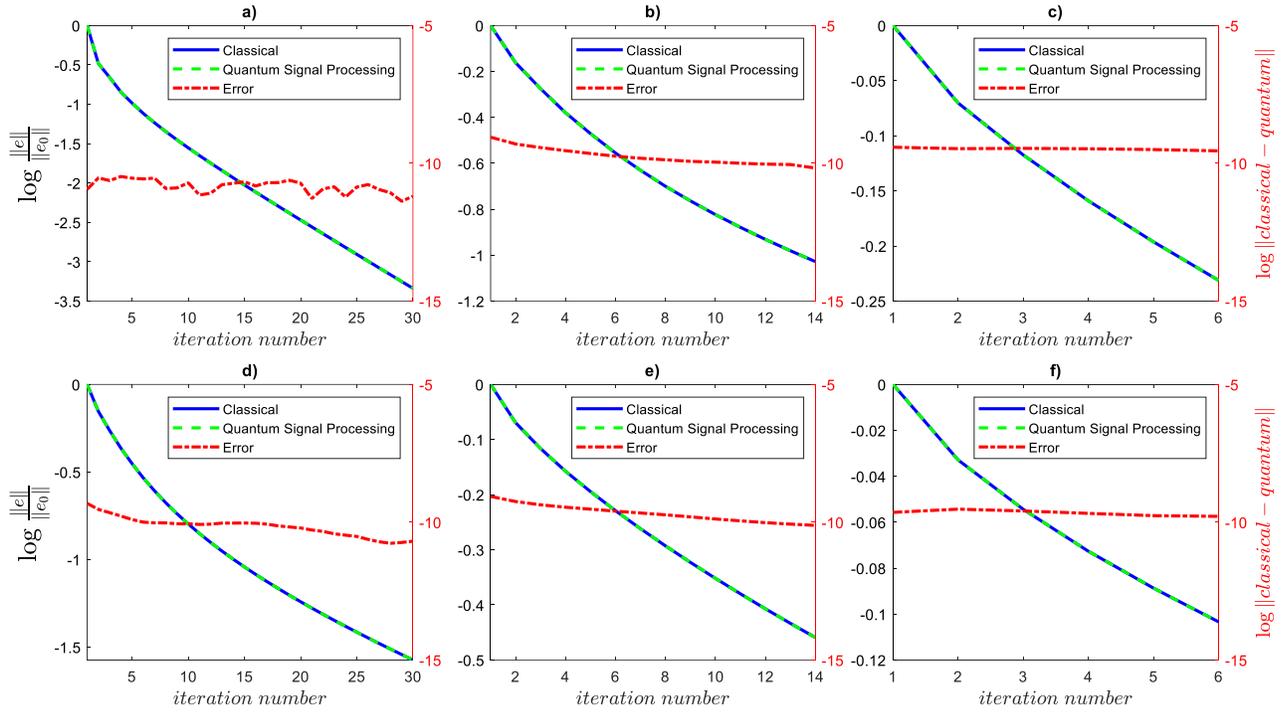

*Figure 2 Error convergence using the quantum signal processing solver using Qiskit's statevector simulator for Case a: a) N=4, b) N=8, c) N=16, Case b: d) N=4, e) N=8, f) N=16.*

In Figure 3 we plot the convergence of stationary linear iterations. The error $\frac{\|x_l - A^{-1}b\|}{\|x_0 - A^{-1}b\|}$ is plotted against $l/\kappa$. The convergence rate is monotonic, and all the convergence rates coincide as expected for stationary linear iterations of positive-definite systems. Note that unlike Figure 2 we do not plot the error between normalized versions of the iterates. Figure 3 shows that further iterations will monotonically lead to a converged solution.

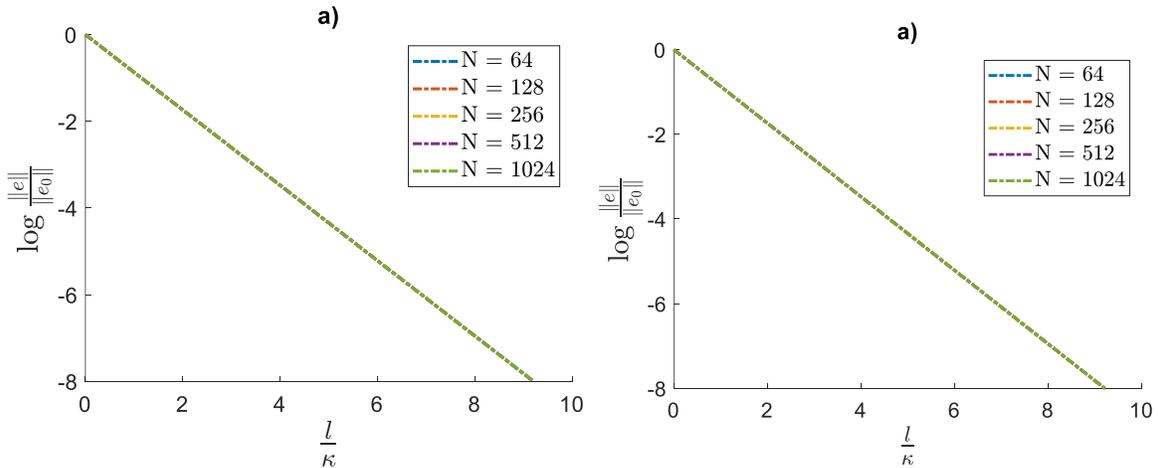

*Figure 3 Classical convergence of Cases (a) and (b) using linear stationary iterations.*

## 4.2. Poisson's problem in 2D

We consider the following Poisson's equation on a square domain:

$Find\ u: \bar{\Omega} \to \mathbb{R}\ such\ that$
$\nabla^2 u - b = 0\ in\ \Omega \in (0, L)$  (46)
$with\ b = 1$

subject to

$u = g\ on\ \Gamma_g$

$-q_i n_i = h\ on\ \Gamma_h$

for the cases listed in Table 2.

*Table 2 Cases a - b considered for problems in 1D*

|        | $\Gamma_g$ | $g$ | $\Gamma_h$ |
|--------|-----------|-----|-----------|
| Case a | $x, y = 0. x = L$ | 0 | $y = L$ |
| Case b | $x, y = 0$ | 0 | $x, y = L$ |
| Case c | $x = 0$ | 0 | $y = 0. x = L$ |

Figure 4 shows the exact solutions of Cases (a)-(c) for grids of $N = 64 \times 64, 64 \times 64$, and $64 \times 65$ nodes respectively.

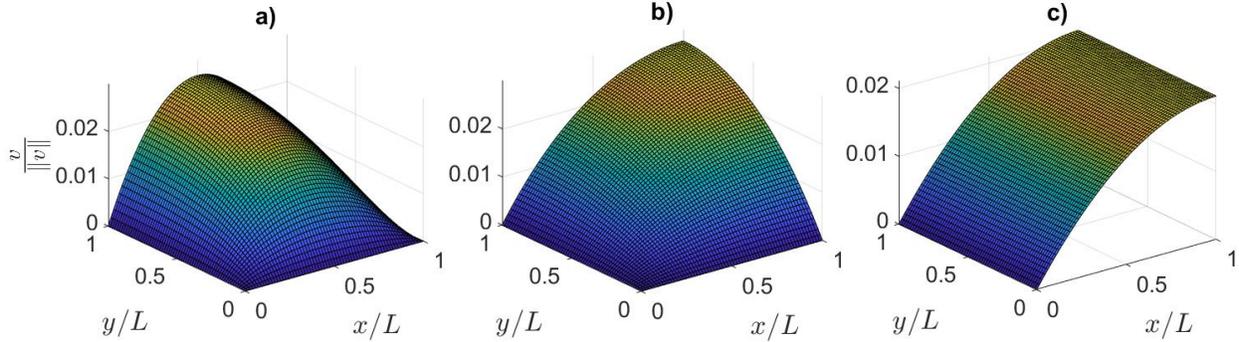

*Figure 4 Exact solutions for Cases (a)-(c).*

Figure 5 shows results of Qiskit's statevector simulation for cases (a)-(c), compared with classical linear stationary iterations. Due to normalization of quantum states, we plot the convergence of the error $\frac{\|e_l\|}{\|e_0\|} = \frac{\left\|\frac{x_l}{\|x_l\|} - \frac{A^{-1}b}{\|A^{-1}b\|}\right\|}{\left\|0 - \frac{A^{-1}b}{\|A^{-1}b\|}\right\|}$. Similar to the 1D cases, we also plot $\left\|\frac{x_{l_{quantum}}}{\|x_{l_{quantum}}\|} - \frac{x_{l_{classical}}}{\|x_{l_{classical}}\|}\right\|$ which shows monotonic convergence. The QSP QLSA precision is chosen as $\delta \leq 10^{-9}$, which is matched by results from the simulator. We solve for various choices of $l$ and $N$ that enable simulation withing the classical memory requirements of Qiskit's quantum statevector simulator.

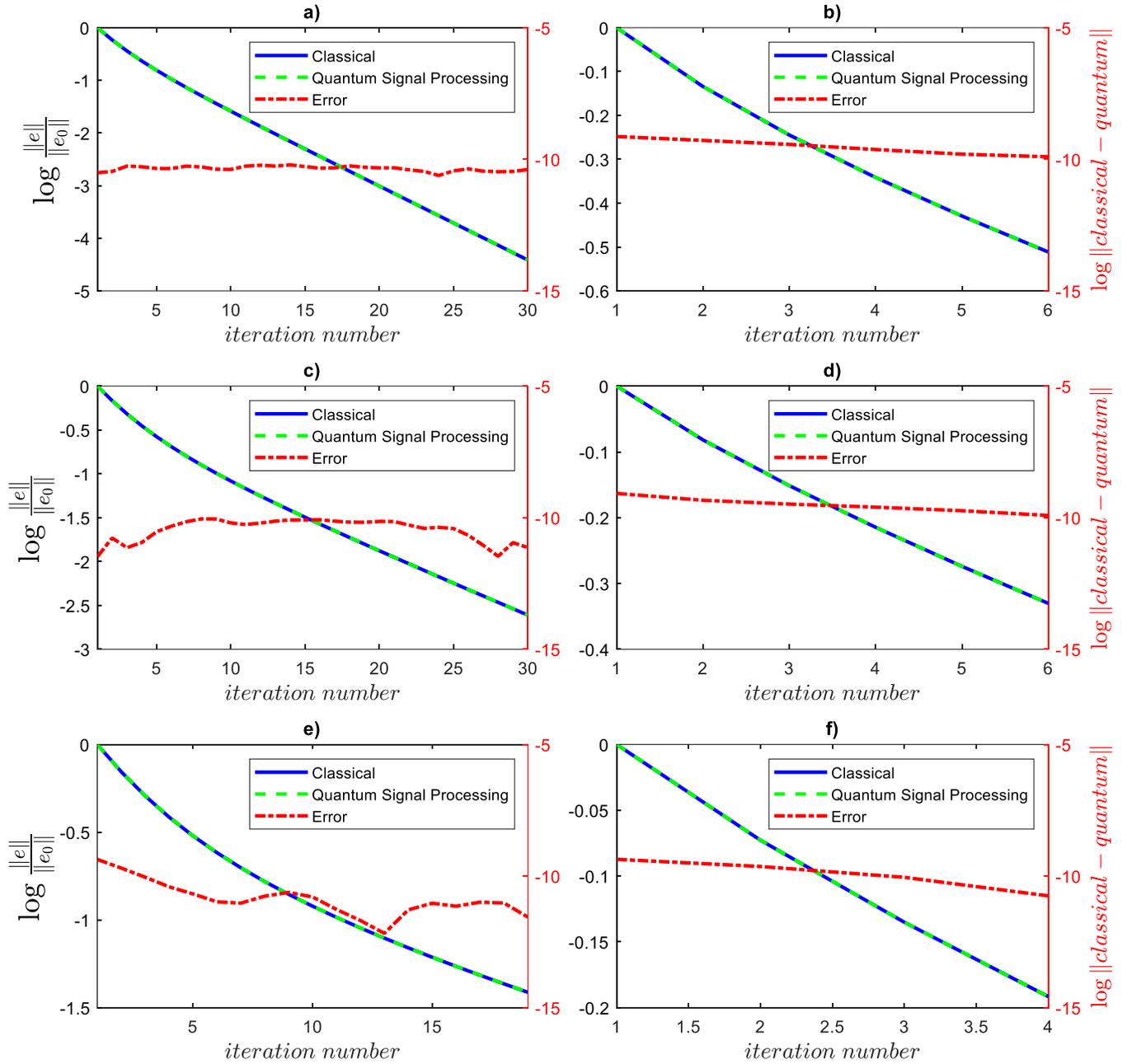

*Figure 5 Error convergence using the quantum signal processing solver using Qiskit's statevector simulator for Case a: a) N=4, b) N=16, Case b: c) N=4, d) N=16, Case c: e) N=6, f) N=20.*

In Figure 6 we plot the convergence of error $\frac{\|x_l - A^{-1}b\|}{\|x_0 - A^{-1}b\|}$ for $l/\kappa$. The convergence is monotonic and constant, and all the convergence rates coincide similar to the 1D cases.

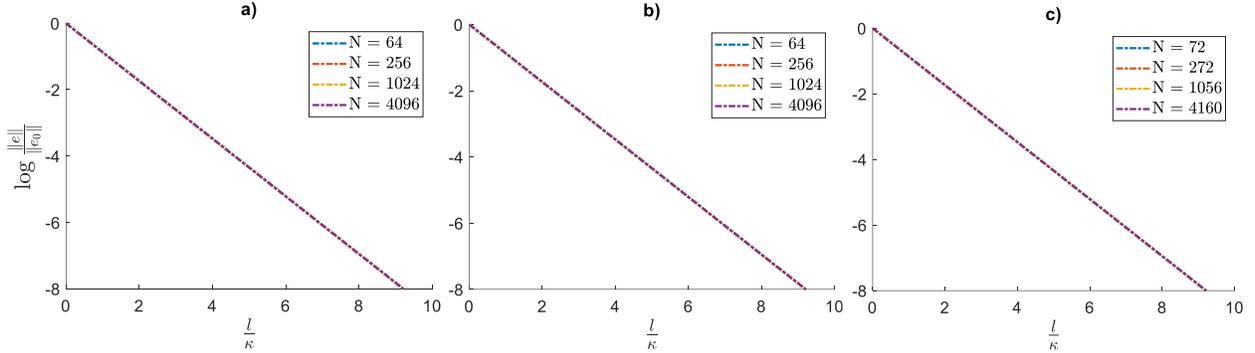

Figure 6 Convergence for Cases (a) – (c) for various N.

## 5. Conclusion

We introduce qRLS, an iterative relaxation method for solving linear systems arising from finite element discretization on gate-based quantum computers. Unlike classical methods, qRLS offers exponential efficiency by scaling logarithmically with system matrix size and linearly with iteration count. The method guarantees convergence for positive-semidefinite systems without the need for diagonal dominance. Nonetheless, the relaxation method suffers from slow convergence for larger problems with increasing condition number.

We demonstrate a success probability $p \geq \frac{1}{2}$ independent of the problem size. Amplitude amplification can boost a success probability $p$ to $O(1)$ at the cost of increasing the complexity of an algorithm by $O(\sqrt{1/p})$, which includes access to the input quantum state. In many cases repeated access to the initial state is unacceptable, and the quantum uniform singular value amplification method can be used to boost the success probability to $\approx O(1-q)$ with an overhead of increasing the algorithm running time by a factor of $O(p/q)$ and reduced precision [32]. In the case of our algorithm, we may also increase the success probability by increasing the number of copies. The success probability may be tailored to the particular use case of the qRLS algorithm.

Various techniques are available to speed up the convergence rate of relaxation techniques. Preconditioning is a widely used method which could be integrated into the iterative scheme. Multigrid and domain decomposition are optimal iterative methods with a classical complexity of $O(N)$ for Laplacian problems. The number of relaxation steps in the multigrid method is $O(1)$, independent of problem size at each grid level, leading to a total requirement of $O(\log N)$ relaxation steps corresponding to $O(\log N)$ grid levels. The classical $O(N)$ complexity in the multigrid method arises from linear operations, sparse matrix-vector products and vector-vector additions, which can be performed efficiently on a quantum computer. The presented block-encoding scheme can be expanded to integrate multigrid linear operations, to possibly reach a complexity of $O(\text{poly} \log (N/\epsilon))$ on quantum computers.

Preconditioning has been proposed in various quantum algorithms to directly use a QLSA on the original system $A$ [15] but explicit implementations of preconditioned QLSAs for general problems are unavailable. However, QLSA's implemented directly on the original system $A$ cannot utilize an initial guess to produce an improved solution. The presented iterative method can also deal with semi-positive-definite systems where $b$ lies in the null space of $A$, unlike QLSA's applied directly [14].

## 6. Acknowledgements

The authors would like to acknowledge the support of this work by NIBIB R01EB0005807, R01EB25241, R01EB033674 and R01EB032820 grants.

## 7. Declarations

The authors have no competing interests to declare that are relevant to the content of this article.